\documentclass[referee]{raa}            % referee version: for submission

\usepackage{graphicx,times}             %for PS/EPS graphics inclusion, new
\usepackage{natbib}
\usepackage{amssymb,amsmath}
\bibpunct{(}{)}{;}{a}{}{,}

\usepackage[a4paper=true,dvipdfm=true,pagebackref=true]{hyperref}
\hypersetup{colorlinks = true, linkcolor = green, anchorcolor = red, citecolor = blue, filecolor = red, pagecolor = red, urlcolor = red}

\begin{document}

   \title{Solar observation with the Fourier transform spectrometer I : Preliminary results of the visible and near-infrared solar spectrum
%\,$^*$
%\footnotetext{$*$ Supported by the National Natural Science Foundation of China.}
}
%   \subtitle{I. Place Your Subtitle Here}

   \volnopage{Vol.0 (20xx) No.0, 000--000}      %%preserved for Editor. DOn't remove!
   \setcounter{page}{1}          %%starting page, preserved for Editor. DOn't remove!

   \author{Xianyong  Bai \inst{1,2} \and Zhiyong Zhang\inst{1,2} \and Zhiwei Feng \inst{1,2} \and Yuanyong Deng \inst{1,2} \and Xingming Bao \inst{1}  \and   Xiao Yang \inst{1} \and Yongliang Song \inst{1} \and Liyue Tong \inst{1,2} \and Shuai Jing \inst{1}
   }
%% Here is an example of three authors come from different institutes.
%% For single author or all the authors from an institute, use "\inst{}" only

   \institute{Key Laboratory of Solar Activity, National Astronomical Observatories, Chinese Academy of Sciences, 20 Datun Road, Beijing 100012, China; {\it xybai@bao.ac.cn}\\
%% Please give the E-mail address of the author, to whom future correspondence and
%% offprint requests will be sent.
        \and
          School of Astronomy and Space Science, University of Chinese Academy of Sciences, 101408,Beijing. \\
        %\and
         % Yunnan Observatories, Chinese Academy of Sciences P.0.Box110, 650011, Kunming, China. \\
\vs\no
   {\small Received~~20xx month day; accepted~~20xx~~month day}}

\abstract{The Fourier transform spectrometer (FTS) is a core instrument for solar observation with high spectral resolution, especially in the infrared. The Infrared System for the Accurate Measurement of Solar Magnetic Field (AIMS), working at 10-13 $\mu m$, will use a FTS to observe the solar spectrum. The Bruker IFS-125HR, which meets the spectral resolution requirement of AIMS but just equips with a point source detector, is employed to carry out preliminary experiment for AIMS. A sun-light feeding experimental system is further developed. Several experiments are taken with them during 2018 and 2019 to observe the solar spectrum in the visible and near infrared wavelength, respectively. We also proposed an inversion method to retrieve the solar spectrum from the observed interferogram and compared it with the standard solar spectrum atlas. Although there is a wavelength limitation due to the present sun-light feeding system, the results in the wavelength band from 0.45-1.0 $\mu m$ and 1.0-2.2 $\mu m$ show a good consistence with the solar spectrum atlas, indicating the validity of our observing configuration, the data analysis method and the potential to work in longer wavelength. The work provided valuable experience for the AIMS not only for the operation of a FTS but also for the development of its scientific data processing software.
%Two small wavelength ranges containing the chromospherical H$\beta$ 486.1 nm line as well as the photospherical Fe I 1.56 $\mu m$ line are selected and compared with the stand solar spectrum atlas published by National Solar Observatory(NSO), USA. The comparison result shows that our observed solar spectrum is consist with the atlas.; The inversion procedure contains zero filling, triangular apodization and phase \textbf{shift} correction.
\keywords{Sun: General --- Methods: Observational ---Instruments: Fourier transform spectrometer }
}

   \authorrunning{X. Y. Bai, Z. Y. Zhang, \& Z. W. Feng, et al. }            %author_head in even pages
   \titlerunning{Solar spectral observation with a FTS }  % title_head in odd pages

   \maketitle
%% The author head (on even pages) and the title head (on odd pages) will be
%% automatically extracted from \author{} and \title{}. Whenever the title is too long,
%% you will be asked to supply a shorter one by inserting either \authorrunning{} or
%% \titlerunning{} before \maketitle. Anyway, you can specify your own heads.
%%
%%
%% Note: In the following text body of your manuscript, please note several differences from
%%       other major journals:
%% (1) \subsection{Please Capitalize the First Letter of Each Notional Word in Subsection Title}
%% (2) Please Capitalize the First Letter of Each Notional Word in all tables' captions

%
%________________________________________________ sections below
%
\section{Introduction}

%\sout{spatisal  aaa}
%\textcolor{red}{text aaa}

Astronomical spectrum provides us an unique opportunity to quantitatively investigate the physical parameters of the observed objects, e.g., chemical composition, temperature, abundance, line-of-sight velocity, pressure and the magnetic field, et al. \citep{Tennyson2011}. Up to now, most of the knowledge we learned about the sun, our nearest star, comes from the spectral observations. With the solar spectropolarimetry data, we are able to derive up to tens of physical parameters and even their variation with optical depth with the help of powerful inversion techniques \citep{del2016,Ai1993,Hector2015}. We can then reconstruct the dynamical three-dimensional solar atmosphere so as to better understand different kinds of quiet or active solar phenomena, such as sunspots, granulation as well as solar flares \citep{Fang2010,Feng2020,Xu2005,Li2017}.

%In 1814, Joseph von Fraunhofer discovered many dark lines in the visible solar spectrum. Since then, breakthrough science was made based on visible solar spectrum with ground based solar telescope \citep{Tennyson2011}. Since the early 1960s, the observation with rocker flights and space-based solar telescopes opened a new era to study solar X-ray and ultraviolet spectrum and the solar corona has been studied in detail with these data \citep{Del2018,Tian2017,Huang2019}. Almost at the same time, infrared solar spectrum has been observed with high spectral resolution from the ground with the McMath$-$Pierce Solar Facility \citep{Pierce1969,Lena1970,Turon1970,Noyes1972,Hall1973,Lena1977}.

Infrared solar spectrum contains lots of scientific advantages relative to the other wavelength. Firstly, it is helpful for the accurate measurements of solar magnetic fields. The ability of a magnetic sensitive line is generally represented by Zeeman sensitivity, which is the ratio of the Zeeman splitting divided by the spectral line width and propotional to $g\lambda$ \citep{Penn2014}. Here g is the Land$\acute{e}$ factor of the selected spectral line and $\lambda$ is the wavelength. If the infrared lines with larger $\lambda$ are used, we can get much higher Zeeman sensitivity \citep{Bruls1995,Solanki2006}. Secondly, many molecular rotation-vibrations lines exist in the infrared waveband and provide unique ways to probe the cool parts of the solar atmosphere, e.g., the well known CO lines near 4.6 microns \citep{Ayres2002,Solanki1994,Uitenbroek1994,Li2020}. Lastly, we can also probe different heights of the solar atmosphere just using continuum radiation because the infrared wavelength covers a wide range from 0.7 to 1000 $\mu m$ \citep{Penn2014}. So most of the new constructed solar telescopes are equipped with the post-focus instrument working in the infrared wavelength, such as the cryogenic infrared spectrograph (CYRA) for the Goode Solar Telescope (GST) and the Cryogenic Near-Infrared
Spectro-Polarimeter for the Daniel K. Inouye Solar Telescope \citep{Cao2010,Rimmele2020}. We are going to usher in a golden age for solar infrared observation in the coming decades.

To accurately measure solar magnetic fields, a new telescope named the Infrared System for the Accurate
Measurement of Solar Magnetic Field (AIMS) is under construction in China. The Mg I 12.32 $\mu m$ line is selected as the working line because it has the largest magnetic sensitivity among our known spectral lines so far. The required spectral resolution is 0.6 \AA \ at 12.32 $\mu m$ \citep{Deng2016}, with a resolution power of 205333. AIMS employs the Fourier transform spectrometer (FTS) to realize the high resolution power. In the 1980s, McMath$-$Pierce Solar Facility at National Solar Observatory, USA, also used a FTS to discover the Mg I 12.32 $\mu m$ line \citep{Brault1972,Brault1978,Chang1983}. In addition, the FTS is employed by many solar spaceborne missions to obtain the middle and far infrared solar spectrum, e.g., the ATMOS (Atmospheric Trace Molecule Spectroscopy) experiment on Spacelab 3, and the ACE-FTS (Atmospheric Chemistry Experiment ) onboard the Canadian SCISAT-1 satellite \citep{Hase2010,Farmer1989,Farmer1994}.
% mainly based on the principle of Michelson interferometer
 %(For a grating spectrograph, there is a limitation of the longest wavelength, which is determined by the grating constant and the spectrum order according to the grating equation. As a result, an Echelle grating and a higher order spectrum are usually used to observe the near infrared. )the Gratings are generally used in the solar spectrograph to obtain the visible and near infrared (IR) (0.7-5 microns) solar spectrum \citep{Penn2014,Iglesias2019,Kuhn2012}.
%If we go to longer wavelength, grating with larger size are needed to achieve the same resolving power.  as an example, its working wavelength is the Mg I 12.32 microns line and the . The required width of the grating is more than one meter, making it extremely difficult to manufacture. The alternative is to employ the , which  In order to meet the spectral resolution of AIMS, the moving mirror of FTS needs to move along a guider rail with a length about one meter, which is relatively easy to implement.

Unfortunately, the FTS has never been used by any solar telescopes in China in the past. Hence one of the main problems encountered by AIMS is how to obtain solar spectrum with a FTS. In the paper, we carried out an experiment aiming to observe the solar spectrum with our newly installed FTS at Huairou Solar Observing Station (HSOS), National Astronomical Observatories of China. The purpose of the experiment is to get the experience about the observing configuration of a FTS as well as its data reduction. The paper is arranged as follows. The principle of the FTS is described in Section 2, along with the brief introduction of our experimental system. The obtained interferograms and their inverted spectrum with our proposed method in the visible and near-IR wavelength are presented in Section 3, followed by the conclusion and future perspective part.

%Different from the raw data taken by a grating spectrometer, the raw data from FTS are the interferogram rather than the spectrum. So more calibration steps are need in the FTS to generate the solar spectrum that can directly be used by the solar physicists. Up to now, we do not know exactly what are the necessary calibration steps that should be used for solar observation.

\section{Principle of the Fourier transform spectrometer and Introduction of our experimental system}

\subsection{Principle of the Fourier transform spectrometer}
\begin{figure}
\includegraphics[width=\textwidth, angle=0]{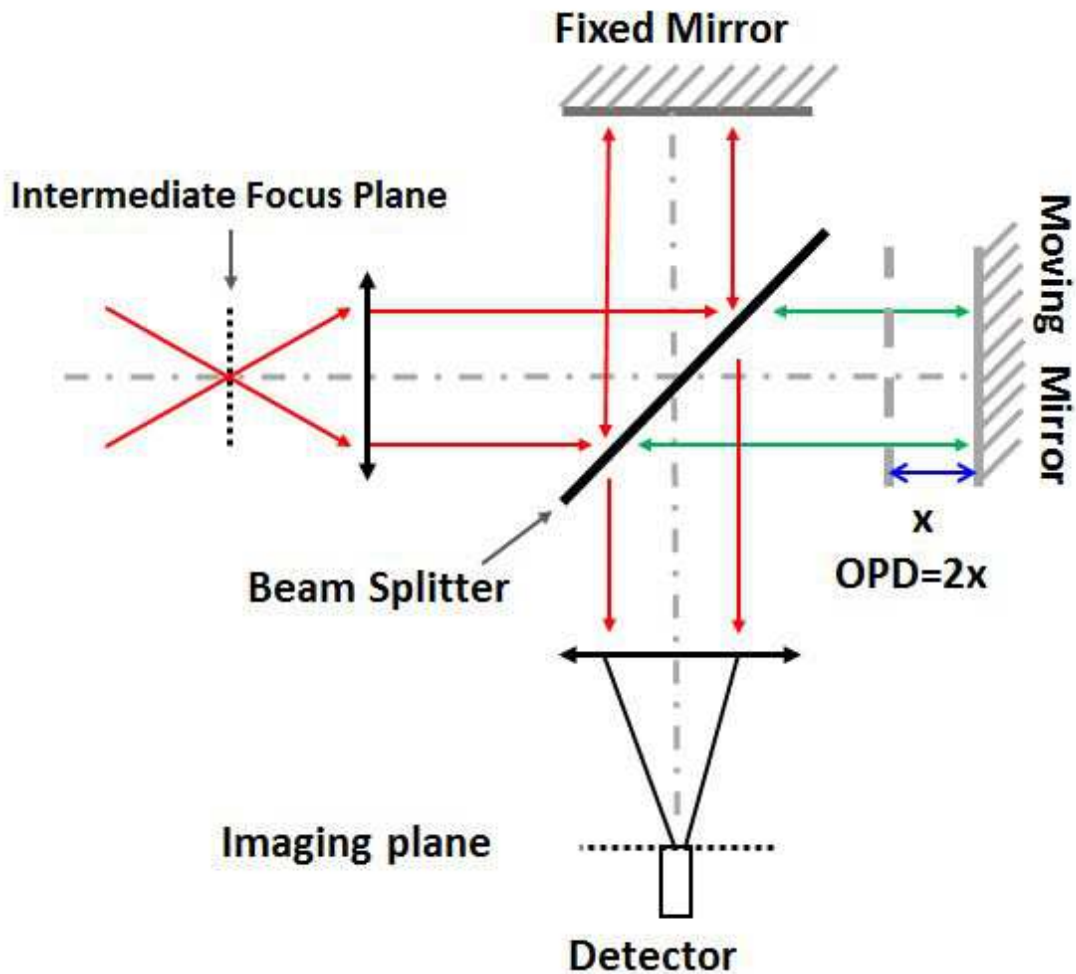} \caption{Schematic diagram of a time-modulated FTS.  }
\label{fig.1}
\end{figure}

 Figure \ref{fig.1} shows the schematic diagram of a time-modulated FTS. The light beam from the solar telescope is firstly focused on the intermediate focal plane and then collimated by a collimating mirror. The collimated light beam further goes through the core part of a FTS, i.e., the Michelson interferometer compartment. The Michelson interferometer generally consists of a beam splitter, a fixed mirror and a moving mirror. The collimated light is divided by the beam splitter into the transmitting and reflecting beams. The transmitting light beam is reflected by the moving mirror and then reflected by the beam splitter (See the green arrows in Figure \ref{fig.1}), which has an optical path of $L_{1}$. The reflecting light beam is reflected by the fixed mirror and then goes through the beam splitter, having an optical path of $L_{2}$. Finally, two beams with different optical paths of $L_{1}$ and $L_{2}$, respectively, recombine into a single beam and interfere with each other. As the moving mirror moves along a precision-steel rod, the optical path $L_{1}$ varies with time. The optical path difference is different at different time, and a series of interference signals is recorded on the detector, forming the so-called interferogram.

For a polychromatic light, the relationship between the interferogram $I(L)$ and the target spectrum $B(\sigma)$ is {\citep{Bates1978,Martin2017,Codding1973}}:

\begin{equation}
I(L)=\int_{-\infty}^{+\infty}B(\sigma) \cos(2\pi\sigma L)d\sigma,
\label{eq1}
\end{equation}

where $\sigma$ represents the wave number ( $\sigma=1/\lambda$, $\lambda$ is the wavelength) and $L$ is the optical path difference (OPD), i.e. $L_{1}$-$L_{2}$. If we want to recover the target spectrum $B(\sigma)$ from Equation \ref{eq1}, the inverse Fourier cosine transform of the interferogram needs to be taken, as shown in Equation \ref{eq2}:
%$B(\sigma)$ represents the solar spectrum over the entire electromagnetic wave here.

\begin{equation}
B(\sigma)=\int_{-\infty}^{+\infty}I(L) \cos(2\pi\sigma L)dL.
\label{eq2}
\end{equation}

From Equation \ref{eq2} and the theory of Fourier transform, the highest spectral resolution of a FTS is determined by its maximum OPD, which depends on the maximum displacement x of the moving mirror (OPD=2x). In theory, the spectral resolution can be infinity. However the longer displacement is, the more difficult to control the moving mirror. Moreover, the volume of a FTS is always finite in reality. So each FTS has a limited spectral resolution $\delta B$, which is proportional to $1/L$. In the case , Equation \ref{eq2} becomes:

 \begin{equation}
B(\sigma)=\int_{-L}^{+L} I(L) \cos(2\pi\sigma L)dL.
\label{eq3}
\end{equation}

The relationship between the limited OPD and the full width at half maximum (FWHM) of the FTS instrument (or $\delta B$) is FWHM $=0.6/L$ because the Fourier transform of a rectangular function (with a width of 2L from -L to L) is a sinc function \textbf{($\sin L/L$)}. The sampling resolution of a FTS is $\frac{1}{2L}$ from Equation \ref{eq3}. Due to the sidelobes or ringing effects of the sinc function, an apodization function is generally used to remove the effect with the sacrifice of reducing spectral resolution. The $\delta B$ becomes 0.9/L for a triangular apodization function \citep{Davis2001}.

For AIMS, the required spectral resolution is 0.6 \AA \ at 12.32 $\mu m$, corresponding to 0.004 $cm^{-1}$. If a triangular apodization function is used, the OPD is $0.9/0.004=225$ cm at least. The minimum displacement d of the moving mirror is 112.5 cm. So we selected a FTS from Bruker Corporation from Germany for our experimental system. Its production model is IFS-125HR, with a maximum OPD of 258 cm. The OPD can be configured from 0 to 258 cm, indicating that we can gather solar spectrum with different spectral resolution. It is worthy to mention that $I(L)$ is a continuous function in equation \ref{eq3}. In reality, we has discrete sampling. Hence the equation can be rewritten below:

\begin{equation}
B(\sigma_j)=\sum_{j=-N}^{N} I(L_n) \cos(2\pi\sigma_j L_n),
\label{eq4}
\end{equation}

where $I(L_n)$ is the interferogram obtained with the OPD of $L_n$ and $B(\sigma_j)$ indicates the real spectrum at the wavenumber $\sigma_j$. The OPD between two adjacent sample interferogram, i.e., $L_{n+1} - L_n$, is the sampling interval $\delta_{opd}$. $N$ is the number of sampling points. The expression of the maximum OPD is : $OPD$$ \ =\  \delta_{opd} \ \times N$.

According to the Nyquist sampling theorem, the largest wavenumber $\sigma_j$ is determined by the sampling interval $\delta_{opd}$ with the relationship of $\sigma_j \le  \frac{1}{2\times \delta_{opd}} $. That is to say, the obtained wavenumber range is from 0 to $ \frac{1}{2\times \delta_{opd}} $ in theory ( or $2\times \delta_{opd}  \le \lambda \ \le  \infty $ with the wavelength unit). For a certain target wavelength, the best $\delta_{opd}$ can be calculated with the above relationship. Generally, one can use a frequency stabilized laser worked in the visible wavelength to ensure equal interval sampling because the Fourier cosine transform of the laser is a cosine function. The laser wavelength $\lambda_{laser}$ of Bruker IFS-125HR FTS is 632 nm and the useable $\delta_{opd}$ is $N\times\frac{\lambda_{laser}}{4}$. For example, the effective wavelength range is from 316 nm to $\infty$ in theory if the $\delta_{opd}$ is set to $\frac{\lambda_{laser}}{4}$. Larger $\delta_{opd}$ is needed for longer wavelength.

From the above mentioned description, we summarized the advantage of a FTS. Firstly, one can observe solar spectrum with the required spectral resolution by setting appropriate OPD value. Secondly, it covers a broad wavelength range ($2\times \delta_{opd} \  \le \lambda \ \le  \infty $) at a single measurement, which is limited by the $\delta_{opd}$, the transmittance or reflectivity of the optical elements and the response range of the detector in reality. Thirdly, as theoretical longest wavelength of a FTS is infinity, it is more suitable for observing solar spectrum at longer wavelength with extremely high spectral resolution, e.g. the middle and far infrared wavelength. The moving mirror of a FTS generally moves along a precise guider rail and can easily move on the order of meters, resulting very large OPD and high resolution. Lastly, the rough wavelength calibration of a FTS is easy due to the equal interval sampling $\delta_{opd}$ from the laser. Once the $\delta_{opd}$ and the number of sampling points are determined, the OPD can be obtained. The wavenumber (wavelength) is also known after the Fourier cosine transform.

To better demonstrate the reason for the AIMS employing a FTS to observe solar spectrum,  we compared a FTS with an Echelle grating spectrograph generally used for high spectral resolution observations in the visible and near-infrared wavelength. The longest wavelength of the Echelle grating spectrograph is determined by the grating constant \textit{d} and the spectrum order \textit{m} according to the grating equation,

\begin{equation}
 2 \times \ d \times\sin \alpha=m \lambda,
\label{eq5}
\end{equation}

where $\alpha $ represents the blazing angle. For example, the largest wavelength occurs at the first spectrum order and equals to 31.645 $\mu m$ for a grating with the $d$ and $\alpha$ of 31.6 $\mu m$ and $71\deg$, respectively. To realize the needed resolution power of 205333 for the AIMS at 12.32 $\mu m$, the required length of the grating is 133.9 cm if the theoretical power of $m\times N_{graing}$ is used, where $N_{graing}$ is the total number of the grating lines. Such a long grating is extremely difficult to manufacture.

\subsection{Brief introduction of our sun-light feeding experimental system and newly installed FTS}

\begin{figure}
\includegraphics[width=\textwidth, angle=0]{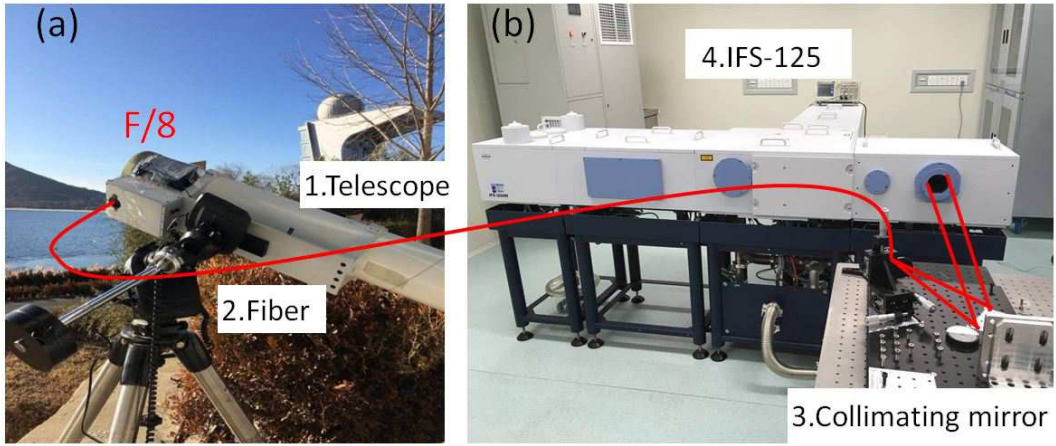} \caption{The experimental system. (a): The equatorial and the Newtonian telescope. (b): The collimating mirror and the Bruker IFS-125HR FTS. The focus of the telescope and the collimating mirror are flexibly linked by the fiber. }
\label{fig.2}
\end{figure}

To get the experiences about the observing configuration of a FTS as well as its data reduction, we employed the Bruker IFS-125HR FTS with a point source detector. The FTS used for AIMS with a detector array of $64 \times 2$ is in development now.  The Equipped parameters of the Bruker IFS-125HR (see Figure \ref{fig.2}) are summarized in Table \ref{tab1}. Its maximal OPD is 258 cm and it has a broad spectral range of 0.4-25 $\mu m$ by selecting different beam splitters and detectors. As a contrast, the maximal OPD of the FTS at McMath$-$Pierce Solar Facility is 100 cm, while its working spectral range is 0.2-20 $\mu m$ \citep{Brault1972,Brault1978}. So the Bruker IFS-125HR FTS has much longer OPD resulting in better spectral resolution in theory. As the spectral range of the fiber used in the current experimental system is 0.275 to 2.1 $\mu m$, only the visible and the near-IR solar spectrum can reach the FTS, which is presented in the following.
\begin{table}
\begin{center}
\caption[]{ Configurations of the Bruker IFS-125HR (new FTS installed at HSOS).}\label{tab1}
\begin{tabular}{clclcl}% 通过添加 | 来表示是否需要绘制竖线
 \noalign{\smallskip}\hline
  wavelength range &  beam splitter & Detector (material, size)\\
 \noalign{\smallskip}\hline
     0.4-1.05 $\mu $m& Dielectric coating on Quartz & Silicon diode, 1 mm$ \times $ 1 mm \\
    0.91-5.4 $\mu $m & Si on CaF2 & InSb detector (cooled with liguid $N_{2}$), 1 mm$ \times $ 1 mm\\
	1.6 -16 $\mu $m  & Ge on KBr& MCT detector(cooled with liguid $N_{2}$), 1 mm$ \times $ 1 mm\\
     2 -25 $\mu $m & Ge on KBr & DLATGS detector, 1 mm$ \times $1 mm\\
 \noalign{\smallskip}\hline
\end{tabular}
\end{center}
\end{table}	

To feed the sunlight from visible to near-IR wavelength into our newly installed FTS at HSOS, we set up a temporary and simple experimental system. It contains a Newtonian reflector, a fiber, a collimating mirror and the FTS, as shown in Figure \ref{fig.2}. The Newtonian reflector with an aperture of 10 cm and the focal ratio of 8, is installed on an equatorial platform to realize the pointing as well as the tracking to the sun. A fiber with the core diameter of 320 $\mu m$ and the numerical aperture of 0.22 is employed to flexibly connect the sunlight from the Newtonian reflector's focus to the focus of the collimating mirror. The sunlight is further collimated by the off-axis parabolic mirror and then enter the FTS. According to the focal ratio of the telescope and the core diameter of the fiber, the field of view of the gathered sunlight is about 82.5 $^{\prime\prime} \times 82.5 ^{\prime\prime}$.

% The IFS 125HR is the ultimate instrument for the high resolution IR spectroscopy due to its outstanding performance in the spectral resolution and broad spectral range. The schematic diagram of the FTS can be found in Figure \ref{fig.2}.
%The second part is the interferometer compartment, being the core part of the FTS.

%The third part of the FTS is the sample compartment, which is generally used to place the sample to analyze its various characteristics. It can also be used for the wavelength calibration as well as the calibration of spectral resolution power of FTS with the help of a absorption cell. When we taken solar observations, there is no sample required in the compartment. The last part of FTS is the detector compartments. It can accommodate up to four internal and two external detectors. In the paper, we only used two types of detectors. One is the visible detector and the other is the near-IR detector, cooled with liquid nitrogen.

%In short, it includes four parts. The first part is a source compartment, which has several built-in light sources and also has the ability to receive the external illuminated light sources, such as the sunlight.

\section{The Observed interferogram and inverted Solar spectrum in the visible and near-infrared wavelength}

In the above section, we introduced the main principle of the FTS and our experimental system. The observed interferogram and its corresponding inverted solar spectrum are arranged in this section, following with a comparison with the solar spectrum atlas obtained by the FTS of National Solar Observatory (NSO), USA. During the observation, we do not use additional narrow band optical filters, so the broadband solar spectrum is obtained.
%三次测量，每次测量重复100次，35分钟一次，共持续 12:02 到13:47，持续105分钟。一次scan 是21 秒。积分30分钟得到这张图.10khz,0.1 cm-1;620000./(4.*10000.)=15.5s, 扫描周期10khz，每个周期采样4次
%0.01 cm-1,40khz; x2.5倍 近红外每次耗时21*2.5=50s
%确定干涉信号及光谱的单位

%zero filling: four points
%apodization:
%Phase correction: Mertze.
%确认一下干涉图和光谱图中的信号是否是电压值（单位mv）
%干涉图
%光谱图

\subsection{The observed solar spectrum with the Bruker IFS-125HR FTS in the visible wavelength}

We firstly try to take test observation in the visible wavelength. During the observation, the FTS is equipped with the Quartz beamsplitter and the silicon diolde detector listed in the first row of Table \ref{tab1}. The target wavelength range is 0.4-1.05 $\mu m$. The needed configurations of FTS are the OPD, $\delta_{opd}$, which are determined by the typical width of the visible solar spectrum and the principle of the FTS shown in Section 2. The $\delta_{opd}$ is set to $\frac{\lambda_{laser}}{4}$ because the selected shortest wavelength is 0.4 $\mu m$. The typical width of solar photospheric lines is about 0.1 {\AA}, corresponding to 0.42 $cm^{-1}$ at 0.48 $\mu m$. The one for the chromospheric lines is wider, which is about 0.3 {\AA} \citep{Moore1966}. So the spectral resolution is set to be 0.1 $cm^{-1}$ here, with a maximal OPD of $0.9/0.1=9\  cm$.

Generally the sampling OPD is from -L to L according to Equation \ref{eq3} and \ref{eq4}. Note that the Fourier cosine transform is an even function, indicating the interferogram between -L to 0 equals to that from 0 to L. Hence the spectrum can be obtained even if we just gather half of the interferogram in theory. Half of the time can be saved in the case but with the same spectral resolution. However, an asymmetrical interferogram will result in phase shifts. A short interferogram from $-L_{1}$ ($L_{1} \le L$) to 0 can be taken then to correct the phase shift \citep{Davis2001}. The Bruker IFS-125HR employs such a configuration. So the interferogram with an OPD from -1 to 9 cm are taken here and the one from -1 to 1 cm is used for the phase shift correction. Considering the required $\delta_{opd}$ is $\frac{\lambda_{laser}}{4}$, the sampling numbers for one interferogram are OPD/$\delta_{opd} \ =\ 6\times10^{5} $.  The scanner velocity of the FTS is set to be 0.632992 $cm/s$. The time taken for one scan is 15.7 s. As our FTS employs a photoconductive detector, the integral time can not be manually configured.

\begin{figure}
\includegraphics[width=\textwidth, angle=0]{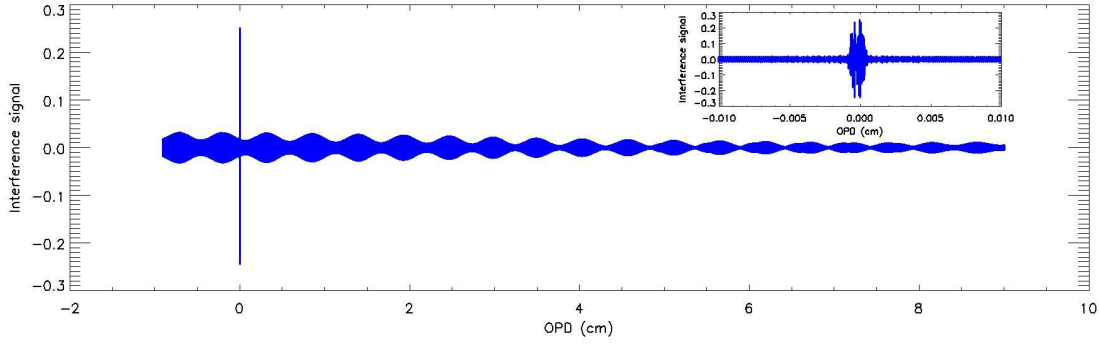} \caption{Original interferogram taken from our experimental system for the visible wavelength. The subplot indicates the double side interferogram with the OPD from -0.01 to 0.01 cm.}
\label{fig3}
\end{figure}

\begin{figure}
\includegraphics[width=\textwidth, angle=0]{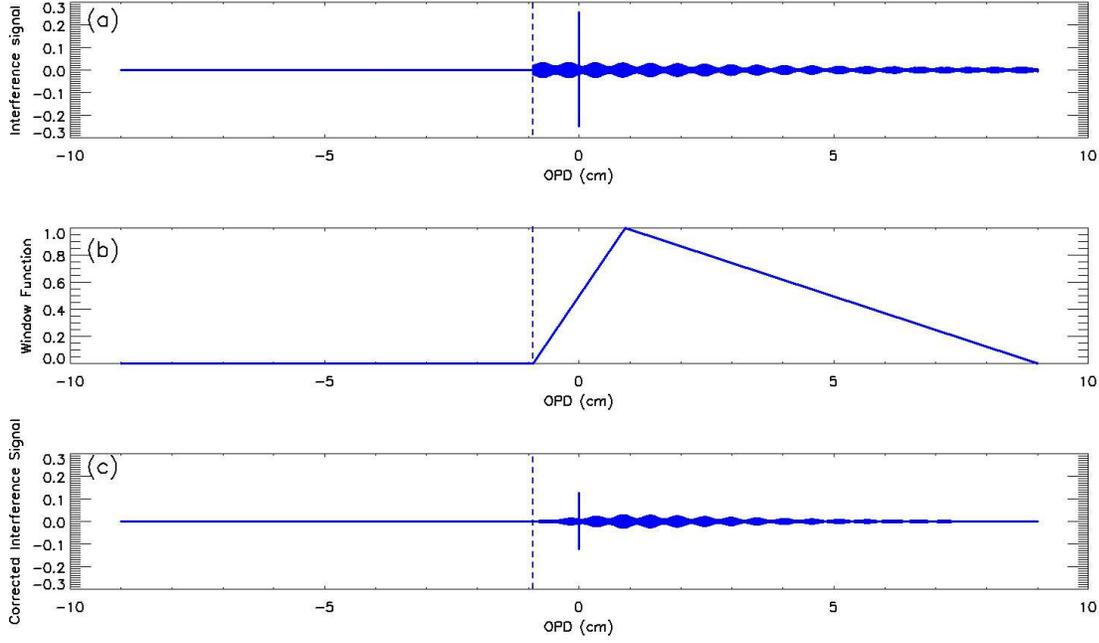} \caption{Panel (a): Interferogram from Figure \ref{fig3} after zero filling. The regions on the left of the dashed line are filled with zero. Panel (b) is the triangular apodizing function. The interferogram after the process of apodization is shown in panel (c).  }
\label{fig4}
\end{figure}

With the above configurations of maximal OPD, $\delta_{opd}$ and the short interferogram used for phase shift correction, we carried out test observation on Dec. 29, 2018, from 04:02 UT to 05:47 UT. Three hundreds of scans are added to improve the signal to noise ratio. The Bruker IFS-125HR can integrate interferogram for many scans. We take one hundred scans for one measurement and three measurements are used. The original unsymmetrical interferogram taken from all the three hundreds scans can be found in Figure \ref{fig3}. The reduction from the interferogram to solar spectrum is listed step by step as follows.

\begin{itemize}
  \item The OPD is extended to the range of -9 to 9 cm and the non-acquired interferogram between -9 and -1 cm is filled with zero. On one hand, we can use equation \ref{eq4} to invert the spectrum as well as its wavenumber value. On the other hand, the sampling resolution of the spectrum can be smaller since we elongated the length of the interferogram. The elongated interferogram after zero filling is arranged in Figure \ref{fig4}a.
  \item Correction of the phase shift. We can rewrite the interferogram in Eq.(6) considering the phase shifts.

\begin{equation}
I(L_n)=\sum_{j=-N}^{N}B(\sigma_j) \exp^{i(2\pi\sigma_j L_n+\phi_j)}=\sum_{j=-N}^{N}B(\sigma_j) \exp^{i\phi_j} \exp^{i2\pi\sigma_j L_n}.
\label{eq6}
\end{equation}

Here $\phi_j$ presents the phase shift, which is nonzero if our sampling grid does not have the point that coincides with zero OPD. The unbalanced dispersion in either arm or the electronics system of the FTS can also introduce phase shifts.

The corresponding inverted spectrum $B(\sigma^1_j)$ is:

\begin{equation}
B(\sigma^1_j)=B(\sigma_j) \exp^{i\phi_j}=\sum_{j=-N}^{N} I(L_n) \times \exp^{-i2\pi\sigma_j x_n}=B_{re}(\sigma_j)+iB_{im}(\sigma_j).
\label{eq7}
\end{equation}

Here $B(\sigma^1)$ is the inverse Fourier transformation of I(L) other than the inverse Fourier cosine transform used in the ideal interferogram in Equation \ref{eq2}. $B_{re}(\sigma_j)$ and $B_{im}(\sigma_j)$ correspond to the real and image components of $B(\sigma^1)$, respectively. Regarding the phase shift $\phi_j$, it can be determined from the double interferogram (Equation 4) with the limited OPD value from -l to 1 cm, according to the following equation:
\begin{equation}
\phi_j=\arctan(B_{im}(\sigma_j),B_{re}(\sigma_j)).
\label{eq8}
\end{equation}

The solar spectrum $B(\sigma_j)$ has only real component. After eliminating the phase shift from the real interferogram, it can be obtained by combining Equations \ref{eq7} and \ref{eq8} \citep{Davis2001}:

\begin{equation}
B(\sigma_j)=Re\{B(\sigma^1_j)\exp^{-i\phi_j}\}.
\label{eq9}
\end{equation}

Another way to correct the phase shift is employing the magnitude of the Fourier transformation. The inverted solar spectrum $B(\sigma_j)$ is :
\begin{equation}
B(\sigma_j)=\sqrt{B^2_{re}(\sigma_j)+B^2_{im}(\sigma_j)}.
\label{eq10}
\end{equation}

\begin{figure}
\includegraphics[width=\textwidth, angle=0]{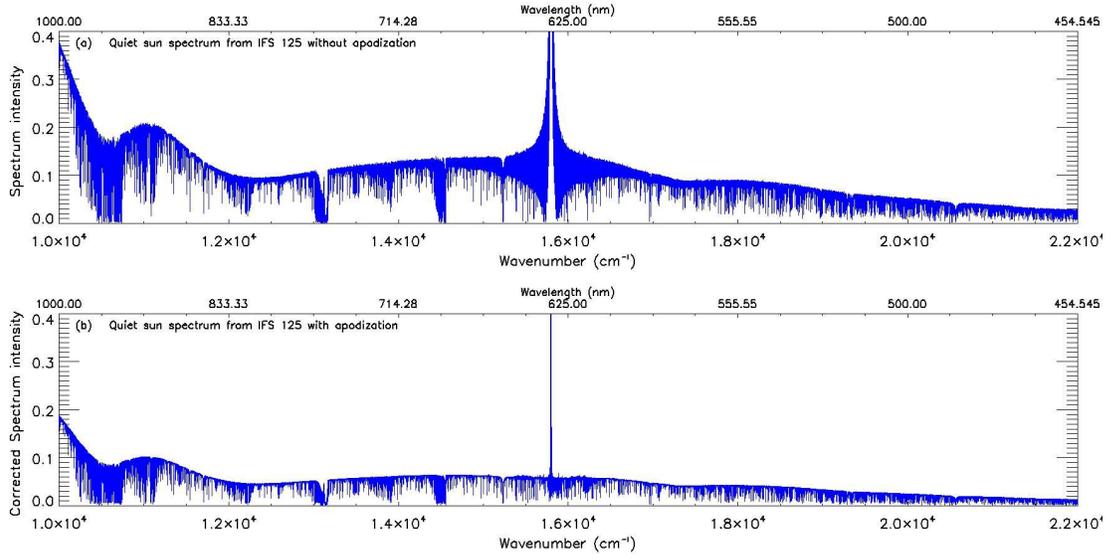} \caption{The solar spectrum inverted from the interferogram in Figure \ref{fig4} in the range of 454.5\ nm (22000 cm$^{-1}$) to 1000\ nm (10000 cm$^{-1}$). The upper and lower panels are the one without and with the triangular apodizing functions.}
\label{fig5}
\end{figure}

In the paper, we use the elongated interferogram in Figure \ref{fig4}a and Equation \ref{eq10} to invert $B(\sigma_j)$. The inverted solar spectrum with the wavenumber selected from 10000 $cm^{-1}$ to 22000 $cm^{-1}$ is presented in Figure \ref{fig5}a. With the known wavenumber from the Fourier transformation, we also obtained the corresponding spectral range, which is from 454.5 nm to 1000 nm. From Figure \ref{fig5}a, we obtained the solar spectrum with a wide wavelength range just from a single interferogram. It is one of the advantage of the FTS. In contrast, tens of scans are needed to cover the same wavelength range for a grating spectrograph because only a narrow waveband is obtained for a single measurement. Moreover, many separate absorption lines exist in the inverted solar spectrum. Most of them are from the Sun while parts of the lines are from the absorption of the earth's atmosphere, which is not identified here. The wider lines in the spectrum correspond to the chromospheric spectral lines due to the higher temperature of the chromosphere, such as the famous $H \alpha$ and $H \beta$ lines with the centeral wavelength of 656.28 nm (15237 cm$^{-1}$) and 486.1 nm (20571 cm$^{-1}$), respectively. The bright line near 632.8 nm (15802 cm$^{-1}$) is the reference laser used for realizing the same sampling interval $\delta_{opd}$.  We also chose a narrow wavelength range to check the profiles of a spectral line. The broad chromospheric $H \beta$ line is selected and shown in Figure \ref{fig6}a. The ringing effect due to the limited length of interferogram can be found.

  \item To remove the ring effect, a triangular apodizing function is employed. We multiplied the interferogram in Figure \ref{fig4}a by a triangular apodizing function seen in Figure \ref{fig4}b. The resulting interferogram is presented in Figure \ref{fig4}c. After the Fourier transformation, we obtained the corresponding solar spectrum with Equation \ref{eq10}. The one from 454.5 nm to 1000 nm can be found in Figure \ref{fig5}b and the selected one near the 486.1 nm is shown with the red line in Figure \ref{fig6}b. Comparing the solar spectrum before and after employing the apodizing function, it clearly illustrates that most of the ring effect disappears.

%  \begin{equation}
%B(\sigma^1_j)=B(\sigma_j) \exp^{i\phi_j}=\sum_{j=-N}^{N} I(x_n) \times APOD \exp^{-i2\pi\sigma_j x_n}=B_{re}(\sigma_j)+iB_{im}(\sigma_j).
%\label{eq11}
%\end{equation}.
\end{itemize}

In order to check the performance of the new installed FTS as well as the quality of our inverson algorithm, we selected parts of the inverted solar spectrum to compare with the solar atlas from the FTS obtained by NSO \citep{Wallace1998}. The broad chromospheric $H\beta$ line is selected with a wavelength range of about 1.2 nm, from 485.4nm (20550 cm$^{-1}$) to 486.6 nm (20550 cm$^{-1}$). Two steps are carried out before the spectrum are used for comparison. The nearby continuum from our inverted solar spectrum is firstly normalized to that from NSO with the green line in Figure \ref{fig6}b. Then the two solar spectrum are registered with each other. Because the wavenumber of the FTS is well determined by the OPD and the $\delta_{opd}$, the shifted value for the registration is very small, which is 0.11 $cm^{-1}$. Comparing our observed solar spectrum near $H\beta$ line (red line) with that from NSO (green line), they agree very well with each other. Both the line depth and the line width of the broader $H\beta$ line as well as the nearby narrow photospherical lines are almost the same. The difference is that our solar spectrum has much lower signal to noise ratio, which can be improved if we integrate a longer time.

\begin{figure}
\includegraphics[width=\textwidth, angle=0]{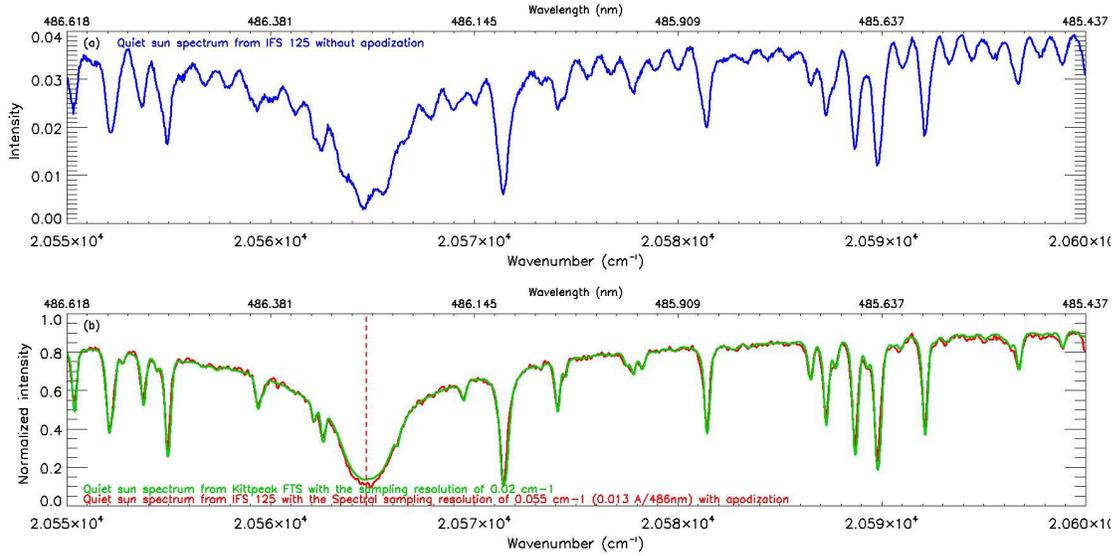} \caption{The blue line in panel a and the red one in panel b are the quiet Sun spectrum near the H$\beta$ line without and with the apodizing function extracted from Figure \ref{fig5}a and \ref{fig5}b, respectively. The continuum of the red one in panel b are normalized with the data taken by the FTS of NSO (green line) for comparison. The dashed line marks the linecenter of the H$\beta$ line. The wavelength used here is vacuum wavelength. So the wavelength 486.1nm in the air corresponds to 486.2 nm in the vacuum.}
\label{fig6}
\end{figure}

%Normalized quiet Sun spectrum (red line) extracted from Figure \ref{fig4} near the H$\beta$ line.
%One range covers the broad chromospheric $H \beta$ line and the other range contains the narrow photospheric $Fe \ I $ 617.3 nm line. The results for the narrower $Fe I $ 617.3 nm line seen in Figure \ref{fig6}is similar. The mean and standard deviation values of the difference spectrum are 0.0043 and 0.0137. The line depth and line width are also nearly the same except that our solar spectrum has lower signal to noise ratio.
%6173 只是一个时刻拍摄的sun_spectrum_.0_181229.10_ff8_bh3terms.fits (6173的不行就删除)
%33分钟，两次。20190101 12:50-13:56；每次重复50scan；每次scan耗时40s;40hz,0.01cm-1
\subsection{The observed solar spectrum with the Bruker IFS-125HR FTS in the near-infrared wavelength}
We also carried out test observations for the wavelength from 1 $\mu m$ (10000 cm$^{-1}$) to 2.2 $\mu m$ (4500 cm$^{-1}$). The longest wavelength 2.2 $\mu m$ is limited by the transmittance of optical fibre. To cover the wavelength, the CaF$_{2}$ beam splitter and the InSb detector are employed, which are listed in the second row of Table \ref{tab1}. Again, we need to configure $\delta_{opd}$, OPD and the short interferogram from $-L_{1}$ to 0 used for phase shift correction. As the shortest wavelength is 1 $\mu m$, the $\delta_{opd}$ is set to be $\lambda_{laser}/2$. One can also use $\lambda_{laser}/4$ in principle at the expense of taking twice as much time as that with $\lambda_{laser}/2$. The typical line width of the photoshperic lines near 1.5 $\mu m$ is about 0.4 {\AA}, corresponding to 0.18$\ cm^{-1}$. The selected OPD is 10 cm here with a spectral resolution of $0.9/10=0.09 \ cm^{-1}$ if a triangular apodizing function is employed according to the principle describe in section 2. $L_{1}$ is set to be -2 cm here. So the interferogram is taken form -2 cm to 10 cm.

\begin{figure}
\includegraphics[width=\textwidth, angle=0]{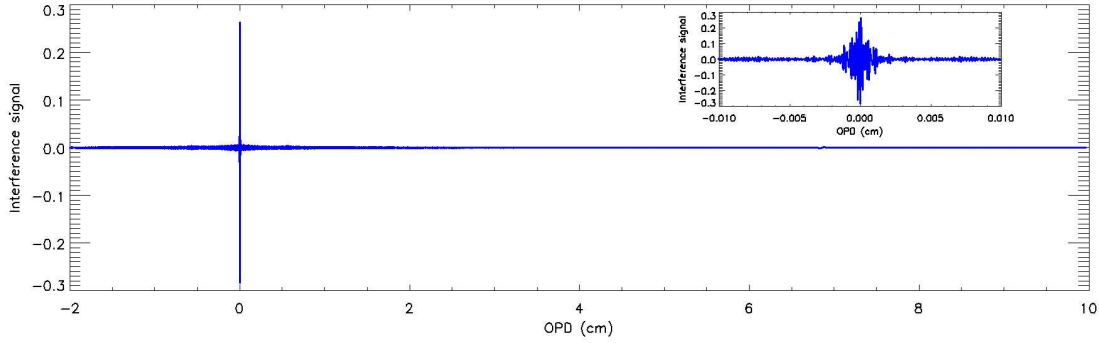} \caption{Original interferogram taken from our experimental system for the near-infrared wavelength. The subplot indicates the double side interferogram with the OPD from -0.01 to 0.01 cm. }
\label{fig7}
\end{figure}

\begin{figure}
\includegraphics[width=\textwidth, angle=0]{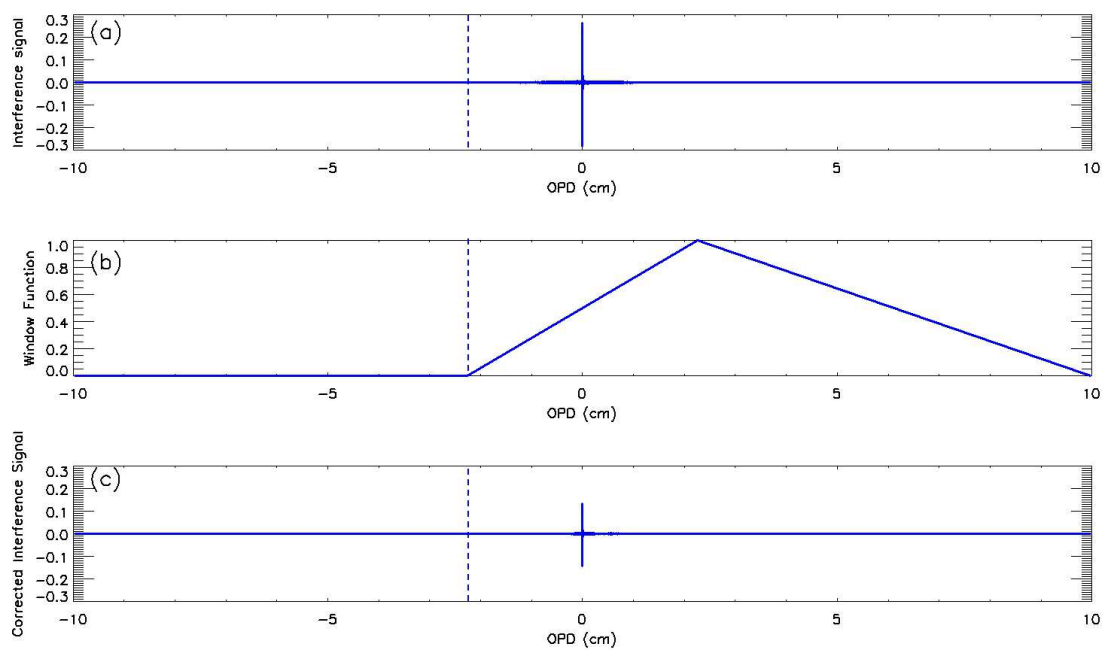} \caption{From the upper to last rows, they are the interferogram from Figure \ref{fig7} with zero filling, the triangular apodizing function and the apodized interferogram, respectively.  }
\label{fig8}
\end{figure}

\begin{figure}
\includegraphics[width=\textwidth, angle=0]{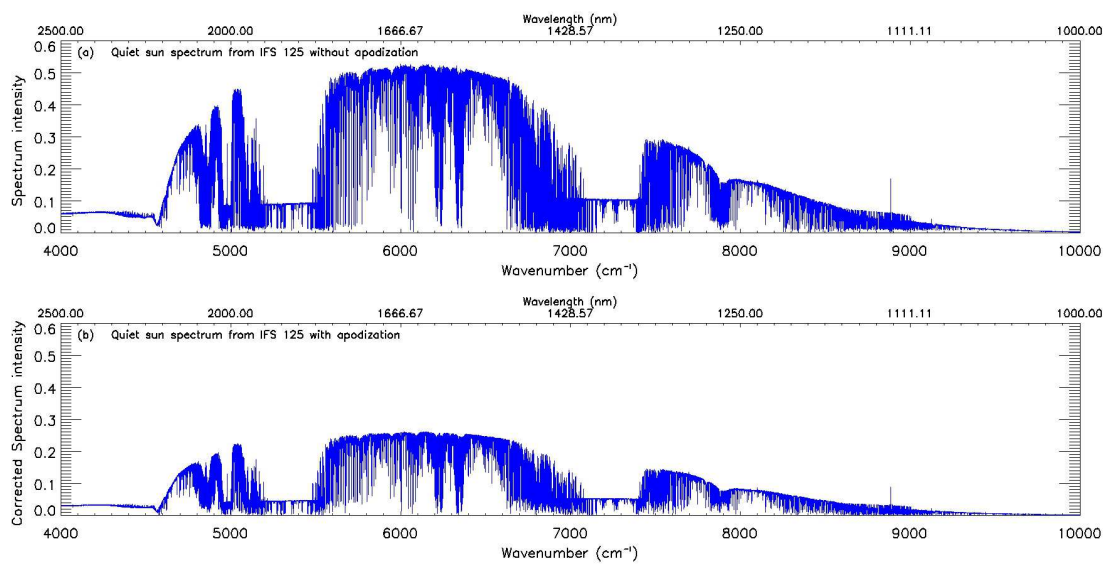} \caption{The solar spectrum inverted from the interferogram in Figure \ref{fig8} in the range of 1000 nm (10000 cm$^{-1}$) to 2222 nm (4500 cm$^{-1}$). The upper and lower rows are those without and with apodizing function. }
\label{fig9}
\end{figure}

The observation was taken on Jan. 01, 2019, from 04:50 UT to 05:56 UT. One hundred scans are added to improve the signal to noise ratio. The original interferogram is arranged in Figure \ref{fig7}. The data reduction of the interferogram is the same as that in the visible wavelength. First, the values of the interferogram with the OPD from -10 cm to -2 cm are filled with zero, as seen from Figure \ref{fig8}a. Then it is multiplied by a triangular apodizing function in Figure \ref{fig8}b. The interferogram after apodization is presented in Figure \ref{fig8}c. We take the Fourier transform of the interferogram with and without apodizing function and the corresponding solar spectrum from 1 $\mu m$ to 2.2 $\mu m$ calculated with Equation \ref{eq10} is shown in Figure \ref{fig9}a and \ref{fig9}b, respectively. We can find that the solar spectrum with apodization has lower intensity. Similar with the spectrum in Figure \ref{fig5}, there are many narrow isolated spectral lines from solar photosphere and some broad spectral lines mainly from the molecular absorption band of the earth atmosphere. For example, the absorption near the wavenumber of 7150 cm$^{-1}$ (1.4 $\mu m$) and 5250 cm$^{-1}$(1.9 $\mu m$) are mainly from the absorption of the water vapour \citep{Hinkle2003}.

\begin{figure}
\includegraphics[width=\textwidth, angle=0]{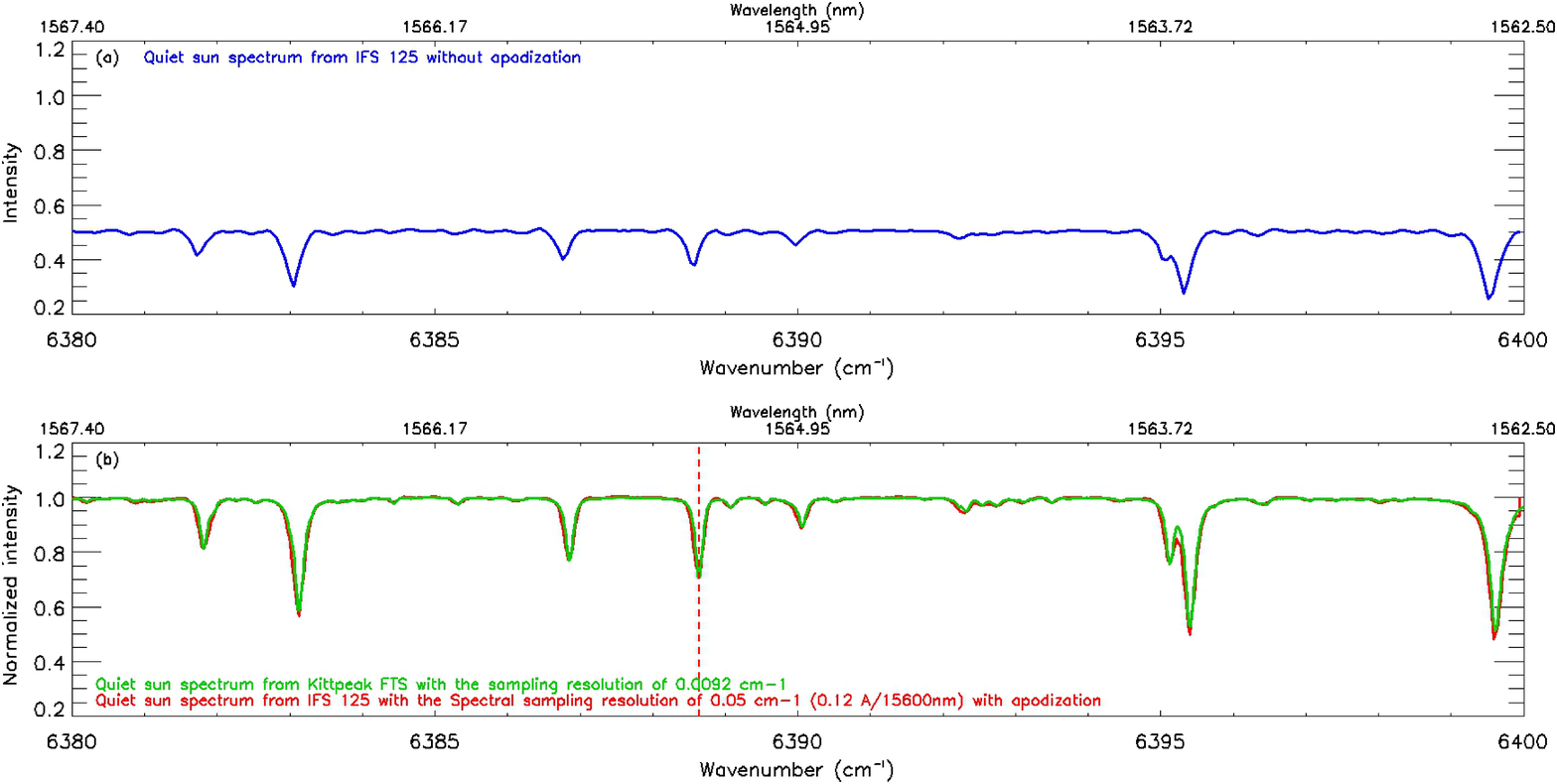} \caption{Similar with Figure \ref{fig6} but for the region near the Fe I 1.56 $\mu m$ line. Panel a and the red line in panel b are the spectrum extracted from Figure \ref{fig9}a and \ref{fig9}b, respectively. The green line are the data taken by the FTS of NSO. The dashed line marks the linecenter of the Fe I 1.5648 $\mu m$ line. The wavelength is also vacuum wavelength.}
\label{fig10}
\end{figure}

At last, we compared our observed solar spectrum with that from the FTS belonging to NSO and the result was shown in Figure \ref{fig10} \citep{Livingston1991}. The selected spectral line is Fe I 1.56 $\mu m$, which is used by many solar telescopes for the accurate measurement of photospherical magnetic field \citep{Collados2012,Cao2010,Liu2014}. The wavelength range is about 4.8 nm.  The continuum is also normalized to that from NSO. The shifted value of the wavenumber is 0.045 $cm^{-1}$ during the registration of the two spectrum. Comparing our inverted solar spectrum before (Figure \ref{fig10}a)and after employing apodizing function (red line in Figure \ref{fig10}b), the ring effect is removed.  From the comparison between the red (our inverted solar spectrum) and the green line that from NSO in Figure \ref{fig10}b, both the line depth and the line width are nearly the same. The signal to noise ratio observed by us is also lower due to the smaller aperature of the light-feeding telescope.

%\begin{figure}
%\includegraphics[width=\textwidth, angle=0]{fig4.4.eps} \caption{The same as Figure \ref{fig5}, but for a narrow absorption lines near the Fe I 617.3 nm. }
%\label{fig6}
%\end{figure}

\section{Conclusion and future perspective}

We installed a Bruker IFS-125HR FTS at Huairou Solar Observing Station with a maximum OPD of 258 cm. The wavelength range is from 0.4 $\mu m$ to 25 $\mu m$, covering the visible, near and mid-infrared wavelength. As a FTS has never used by any solar telescopes in china, we established a temporary sunlight feeding system and tried to gather experience both for the observing configuration as well as the data inversion from interferogram to solar spectrum. The work shown here is useful for the data reduction of the AIMS telescope, which also uses a FTS and will obtain its first light near 2022.
%The experimental system consists of a Newtonian reflector, a fiber and a collimating mirror. The Newtonian reflector is mounted on the equatorial platform so as to realize the pointing and tracking of the Sun. With the help of the system, the solar spectrum in the quiet Sun region is observed with the FTS.

We firstly introduced the principle of a time-modulated FTS and showed that it is more suitable for realizing the spectral resolution of the AIMS. We summarized the main advantages of a FTS. Firstly, it is more suitable for the longer wavelength, e.g., the middle and far infrared waveband. Secondly, it is easy to reach very high spectral-resolution and the resolution can be set optionally by the user. Thirdly, it has broad wavelength range. Lastly, a FTS can give the wavelength value because it has a frequency stabilized laser resulting the well known OPD. The necessary observing configurations needed for a FTS when taking solar spectrum are the appropriate $\delta_{opd}$, maximal OPD as well as the OPD value used for phase shift correction. Also, the directly observed quantity of a FTS is the interferogram. So the Fourier transformation must be employed to recover the solar spectrum.

We carried out test observations with our experimental system in the visible and near-infrared wavelength. According to the target wavelength range and the typical width of the solar spectral lines, we determined the suitable $\delta_{opd}$, maximal OPD as well as the OPD value used for phase shift correction. The interferogram is obtained then. Considering the asymmetric interferogram, infinite OPD and the phase shift in our interferogram, we firstly filled the zero value to make a symmetric interferogram with the OPD from -L to L. Then a triangular apodizing function is multiplied by the interferogram to reduce the ring effect. The final spectrum is the magnitude from the Fourier transformation of the interferogram to correct the phase shift.

With the Bruker IFS-125HR FTS, we successfully obtained the broadband solar spectrum from 0.45 $\mu m$ to 2.2 $\mu m$. Two common used spectral lines, i.e., chromospherical  $H \beta$ 486.1 nm in the visible and photospherical Fe I 1.56 $\mu m$ lines in the near infrared, are compared with those taken from the FTS by NSO. Both the line depth and width are almost the same except a relatively lower signal to noise ratio in the solar spectrum taken by our FTS. The comparison results confirm the effectiveness of our observing configuration and data reduction method.
%Due to the high resolution of the FTS, the spectrum obtained in the paper can also be used for estimating the instrument performance of the other solar spectrograph or narrow-band filter, e.g., its spectral resolution and stray light, which is used by paper (me and xuzhi).

We would like to emphasize that the results shown here just focus on the visible and near-infrared wavelength range mainly due to the limited transmittance profile of the optical fiber. The line width of the solar spectrum is not narrow enough to calibrate the practical spectral resolution of the FTS, indicating we need to find other calibration methods. The data reduction from interferogram to solar spectrum is a preliminary result as well. What is the best apodizing function for the inversion of a solar spectrum? What decides the OPD to derive the phase shift? Is it possible to use Equation \ref{eq9} to correct the phase shift? What is the relationship between the scan time and the signal to noise ratio? These questions need to be further investigated. Moreover, the maximal OPD used in the paper is only 10 cm, much less than the maximal OPD of 258 cm. If the solar spectrum is observed in the mid-infrared, e.g., the wavelength range of 10-13 $\mu m$ selected by the AIMS, we need longer OPD. It's also well known that many spectral lines from earth atmosphere exist in the mid-infrared \citep{Hinkle2003}. The radiation from the instrument, the nearby background and the earth atmosphere contribute a lot in the mid-infrared wavelength. So we need to identify and find a method to correct them. Based on the above considerations, an upgraded experimental system with all reflected mirrors is under construction at HSOS by now. The diameter of its primary mirror is 60 cm. The solar spectrum from 2.2 to 25 $\mu m$ with more photons can be taken then, which is helpful for addressing the above remaining problems.
 %
 %is difficult because the width of the solar lines is is The diameter of its primary mirror is 60 cm. With its help, we can develop and test the methods to remove the influence of the instrumental, background radiation and earth atmosphere. The solar spectrum from 2.2 to 25 $\mu m$ with more solar photons can be taken then. $The difference of FTS used here with AIMS$
%The radiometric calibration is not done by now.
%the non-linear response of the detector and the influence of background radiation from the environment or the instrument will affect the observed solar spectrum.

%\begin{figure}
%\includegraphics[width=\textwidth, angle=0]{fig_last.eps} \caption{Schematic diagram of Huairou Solar Spectrum Telescope. }
%\label{fig.last}
%\end{figure}

\begin{acknowledgements}
We sincerely thank the referee for helpful suggestions that greatly improved the manuscript. We are also grateful for Prof. Kaifan Ji and Song Feng for the discussions on the theory of Fourier transform. This research work is supported by the Grants: 11873062, 11427901, 11673038, 11803002, 11973056, 11973061, 12003051, 12073040, XDA15320102, and XDA15052200.
\end{acknowledgements}

%\begin{thebibliography}{99}
%% you can type \apj for ApJ, \aap for A&A, \apss for Ap&SS, etc. Please consult
%% the macro chjaa.cls. You can also find them in aasguide.tex (AASTeX for ApJ, AJ, PASP)
%% Please follow the format of ChJAA's reference list

%\bibitem[{{Evans}(1949)}]{Evans1949}
%{Evans}, J.~W. 1949, Journal of the Optical Society of America (1917-1983), 39,
%  229

%\bibitem[{{Title} {et~al.}(1976){Title}, {Pope}, {Ramsey}, \&
%  {Schoolman}}]{Title1976}
%{Title}, A.~M., {Pope}, T.~P., {Ramsey}, H.~E., \& {Schoolman}, S.~A. 1976,
%  {Development of birefringent filters for spaceflight}, Tech. rep.

%\end{thebibliography}
\bibliographystyle{raa}
\bibliography{ms}
\label{lastpage}

\end{document}